Article

# Local and Regional Contributions to Tropospheric Ozone Concentrations

Callum E. Flowerday [1], Ryan Thalman [2] and Jaron C. Hansen [1,*]

1. Department of Chemistry and Biochemistry, Brigham Young University, Provo, UT 84602, USA
2. Department of Chemistry, Snow College, Richfield, UT 84701, USA
* Correspondence: jhansen@chem.byu.edu

**Abstract:** The Wasatch Front in Utah, USA is currently a non-attainment area for ozone according to the Environmental Protection Agency's (EPA) National Ambient Air Quality Standards (NAAQS). Nitrogen oxides ($NO_x = NO_2 + NO$) and volatile organic compounds (VOCs) in the presence of sunlight lead to ozone formation in the troposphere. When the rate of oxidant production, defined as the sum of $O_3$ and $NO_2$, is faster than the rate of $NO_x$ production, a region is said to be $NO_x$-limited and ozone formation will be limited by the concentration of NOx species in the region. The inverse of this situation makes the region VOC-limited. Knowing if a region is $NO_x$-limited or VOC-limited can aid in generating effective mitigation strategies. Understanding the background or regional contributions to ozone in a region, whether it be from the transport of precursors or of ozone, provides information about the lower limit for ozone concentrations that a region can obtain with regulation of local precursors. In this paper, measured oxidant and $NO_x$ concentrations are analyzed from 14 counties in the state of Utah to calculate the regional and local contributions to ozone for each region. This analysis is used to determine the nature of the atmosphere in each county by determining if the region is VOC- or $NO_x$-limited. Furthermore, this analysis is performed for each county for the years 2012 and 2022 to determine if there has been a change in the oxidative nature and quantify the regional and local contributions to ozone over a 10-year period. All studied counties—except for Washington County—in Utah were found to be VOC-limited in 2012. This shifted in 2022 to most counties being either in a transitional state or being $NO_x$-limited. Local contributions to ozone increased in two major counties, Cache and Salt Lake Counties, but decreased in Carbon, Davis, Duchesne, Uinta, Utah, Washington, and Weber Counties. Generally, the regional contributions to oxidant concentrations decreased across the state. A summertime spike in both regional and local contributions to oxidants was seen. Smoke from wildfires was seen to increase the regional contributions to oxidants and shift the local regime to be more $NO_x$-limited.

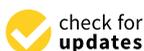

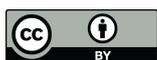

## 1. Introduction

Ozone ($O_3$) is a trace gas present in the Earth's atmosphere. While it plays a role in protecting the planet from harmful ultraviolet radiation from the sun when in the stratosphere, high concentrations of ozone near the Earth's surface, referred to as ground-level (or tropospheric) ozone, are a criteria pollutant and harmful to human health [1,2]. Up to 90% of the Earth's ozone exists in the stratosphere and is naturally occurring [3]. This is commonly known as the "ozone layer". The rest of the Earth's ozone can be found in the troposphere. A fraction of this ozone is naturally occurring, but concentrations above background levels, usually formed from the photochemical reaction of nitrogen oxides ($NO_x = NO + NO_2$) with volatile organic compounds (VOCs) in the presence of sunlight, is problematic [4,5]. $NO_x$ and VOC compounds are emitted into the atmosphere where they react in the presence of photochemically produced radicals, such as OH, to form ozone.

Long-range transport of NO$_x$, VOCs, and even O$_3$ can result in elevated concentrations of ozone in regions where local NO$_x$ and VOC emissions may not be indicative of the overall ozone concentrations measured in the area [6–9].

The total concentration of ozone in an area is the sum of the regional and local contributions. Regional contributions are emissions that are transported over long distances to areas downwind from the sources [6–9]. Regional contributions define the background concentration of ozone in a local region as this baseline concentration of ozone would exist in spite of local contributions to ozone. Varying amounts of regional contributions to ozone would correspond to varying amounts of background, or baseline, ozone. Local contributions refer to the impact on ozone concentrations from emissions from point sources in that specific area. These are sources of primary pollution. These emissions can react with other pollutants in the atmosphere to form ozone in a region. The combined regional and local contributions to ozone concentrations can result in challenging air quality conditions that require a comprehensive approach to mitigate the impact of emissions on air quality by a regulatory authority. The ability to deconvolve the effects of local and regional contributions to a regions' total ozone concentration can help guide mitigation strategies. The creation of ozone isopleths, using speciated VOCs and NO$_x$, can be very costly to construct. These projects require a lot of money, time, and specialized instruments that are not commonly deployed at EPA and state air sampling sites. This study uses current EPA and infrastructure to determine the oxidative state of a region—VOC-limited or NO$_x$-limited—which saves time and money. This information can be used by stakeholders, legislature, and policy makers to make informed decisions about ozone mitigation strategies in a region.

The Wasatch Front, a metropolitan area with an approximate population of 2.5 million, in the north-central part of Utah, USA, and the Uinta Basin, located in eastern Utah with an approximate population of 20,000, is an industrial region with approximately 15,000 oil and gas wells that were designated as "marginal non-attainment areas" for ozone by the Environmental Protection Agency (EPA) in 2018 [10–22]. Even though the ozone standard was made more stringent in 2015 from 75 ppb to 70 ppb, it wasn't until 2018 that the Wasatch Front met or exceeded the NAAQS. In the Wasatch Front, ozone concentrations have been historically driven from a variety of sources, including transportation, industry, and residential emissions. Steps have been taken to reduce ozone concentrations in this area, including stricter emission controls on vehicles and industrial facilities and promoting alternative transportation options, such as public transportation and carpooling. However, these measures have not been enough to reduce the ozone concentrations in the Wasatch Front. This study seeks to understand the regional and local contributions to oxidizers, primarily ozone, in the state of Utah and to identify the limiting reagent, either VOCs or NO$_x$, that could be controlled to lower the concentrations of ozone in the region.

A few previous studies have attempted to measure a full suite of VOCs, nitrogen oxides, ozone, meteorological parameters, and other inputs necessary for photochemical modeling in an effort to better understand the oxidative state of the atmosphere; however, these studies have been limited to modeling only a few regions in Utah. Here, the method of Clapp and Jenkin is applied to assess the oxidative state of 14 counties in the state of Utah [23].

## 2. Method

*2.1. Data Analysis Method*

Clapp et al. described the relationship between ambient levels of O$_3$, NO$_2$, and NO and a method for determining the local and regional contributions to oxidizers, which is defined in their work as the sum of O$_3$ and NO$_2$ [23]. The relationship between O$_3$, NO$_2$, and NO is described as a steady-state interconversion of NO to NO$_2$ and NO$_2$ to NO, as follows [24]:

$$NO + O_3 \rightarrow NO_2 + O_2 \tag{1}$$

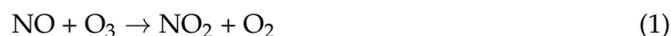

$$NO_2 + h\nu(+O_2) \rightarrow NO + O_3 \qquad (2)$$

Fundamental to more elaborate atmospheric models, such as the Master Chemical Mechanism (MCM), Community Multiscale Air Quality Modeling System (CMAQ), and Comprehensive Air Quality Model with extensions (CAMx), is the mechanism described by reactions 1 and 2 for the formation of $O_3$ [25–27]. The relationship described above becomes non-linear when nitric acid is formed, removing $NO_x$ from the mechanism described by reactions 1 and 2; however, for the current analysis, a linear relationship is assumed between oxidants and $NO_x$ [28]. This photostationary relationship is often referred to as the Leighton Relationship [24,29]. The linear relationship between oxidants and $NO_x$ in this cycle has been shown by Clapp et al. [23].

The data provided, and quality assured for the EPA, by the Utah Division of Air Quality (UDAQ) and the authors include measured $O_3$, NO, and $NO_2$ concentrations from 14 sites located in 14 counties in discrete integral ppb measurements. The authors used a Thermo TECO 49c to measure ozone and a Thermo 42 to measure $NO_x$ in Sevier and a Thermo 49i to measure ozone and a Thermo 42i-TLC to measure $NO_x$. All instruments were zeroed and calibrated using calibration gas standards. These data were used to generate plots of the concentration of oxidizers against the concentration of $NO_x$, where oxidizers are defined as the sum of measured $O_3$ and $NO_2$ concentrations. An example of this can be seen in Figure 1 for air quality data collected for the year 2017 from downtown Los Angeles, CA, USA, North Main Street EPA sampling site (site ID 060371103). The intercept of the linear regression shows the regional contribution to ozone for an area (when $NO_x$ is zero the oxidants are comprised only of ozone as no $NO_2$ is present). The slope of the regression line provides insight into the local contributions to oxidants and discloses if the region is VOC- or $NO_x$-limited. A positive slope shows that the oxidant concentration is increasing with an increase in $NO_x$ and can be interpreted to mean that the area is $NO_x$-limited. Conversely, a negative slope shows that the oxidant concentration is decreasing with increasing $NO_x$ and indicates that the area is VOC-limited.

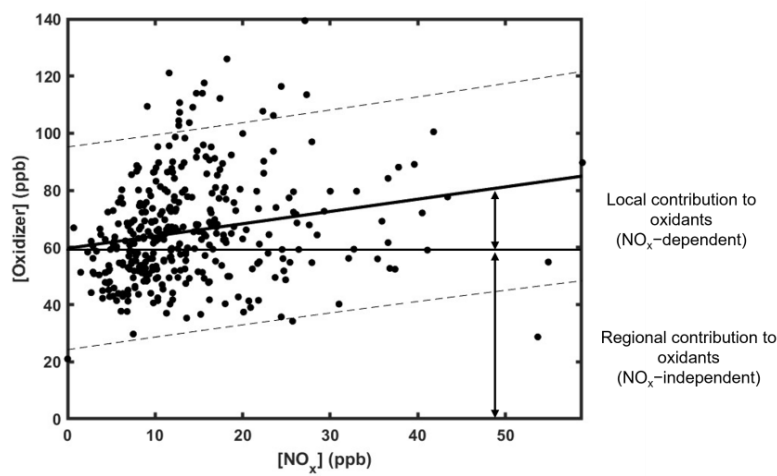

**Figure 1.** Concept graph using data from the Los Angeles, CA, USA, North Main Street EPA sampling site to illustrate how these data are interpreted to deconvolve the regional and local contributions to the overall oxidant concentrations. Dashed lines show the 95% confidence interval for the slope and intercept.

The positive slope in Figure 1 indicates the region is $NO_x$-limited and the intercept shows a background or regional contribution of ozone of 60 ppb.

*2.2. Description of Data*

Ozone, NO, and $NO_2$ concentrations from 14 air sampling sites across 14 counties in Utah, USA were analyzed. These counties, and their respective sampling sites, are listed in Table 1, which also shows their latitude and longitude and the years that data are available from each. Twelve counties were analyzed using data provided by the EPA and two counties were studied using data collected by the authors. Where possible, sites were analyzed using data collected during the years 2012 and 2022 to identify if there has been a change in either the regional or local contributions to the concentration of oxidizers in a region. The data were sorted by using the recorded ozone daily maximum between 7 a.m. and 7 p.m., when the sun was up to drive the cycle described in reactions 1 and 2, and the corresponding $NO_x$ values at that time. This analysis was done by performing a linear regression analysis of oxidizers against total $NO_x$ concentrations. This linear regression was done to determine if each county was $NO_x$-saturated or $NO_x$-limited and to calculate the regional contributions to ozone using the method described by Clapp and Jenkin. When the oxidant concentration is higher than the $NO_x$ concentration, a region is said to be $NO_x$-limited and ozone formation will be limited by the concentration of $NO_x$ species in the region. When the oxidant concentration is lower than the $NO_x$ concentration, a region is said to be $NO_x$-saturated and ozone formation will be limited by the concentration of VOCs in the region. Understanding the contribution to total ozone in a region requires understanding the regional contribution, which is influenced by the prevailing wind direction. This is discussed further in Section 3 of this paper.

**Table 1.** Site details for each of the counties analyzed.

| County | 2022 Site ID | Latitude (°) | Longitude (°) | Altitude (m) | 2012 Site ID |
|---|---|---|---|---|---|
| Box Elder | 490037001 | 41.945874 | −112.233973 | 1372 | |
| Cache | 490050007 | 41.842649 | −111.852199 | 1379 | 490050004 |
| Carbon | 490071003 | 39.609960 | −110.800749 | 1740 | 490071003 |
| Davis | 490110004 | 40.902967 | −111.884467 | 1309 | 490110004 |
| Duchesne | 490130002 | 40.294178 | −110.009732 | 1587 | 490130002 |
| Iron | 490210005 | 37.747430 | −113.055525 | 1690 | |
| Salt Lake | 490353006 | 40.736389 | −111.872222 | 1306 | 490353006 |
| Sanpete | EPH [1] | 39.362074 | −111.580043 | 1695 | |
| Sevier | RCH [1] | 38.766727 | −112.096879 | 1641 | |
| Tooele | 490450004 | 40.600550 | −112.355780 | 1511 | |
| Uinta | 490462002 | 40.206291 | −109.353932 | 1668 | 490462002 |
| Utah | 490494001 | 40.341389 | −111.713611 | 1442 | 490490002 |
| Washington | 490530007 | 37.179125 | −113.305096 | 992 | 490530007 |
| Weber | 490571003 | 41.303614 | −111.987871 | 1317 | 490570002 |

[1] Sampling site operated by the authors.

## 3. Results and Discussion

*3.1. Linear Regression Analysis of Oxidizer Concentrations against $NO_x$ Concentrations*

3.1.1. Annual Data for Each County

Figures 2–15 plot the daily maximum in oxidant and $NO_x$ concentrations for each of the 14 sampling sites analyzed in the years 2012 and 2022. A positive slope, as observed in Figure 2 (Box Elder County 2022), Figure 4B (Carbon County 2022), Figure 6B (Duchesne County 2022), Figure 12B (Uinta County 2022), and Figure 14A (Washington County 2022), indicates that the county is $NO_x$-limited. Negative slopes, observed in Figure 3A,B (Cache County), Figure 3A (Carbon County), Figure 5A,B (Davis County), Figure 6A (Duchesne County), Figure 7 (Iron County), Figure 8A,B (Salt Lake County), Figure 9 (Sanpete County), Figure 10 (Sevier County), Figure 11 (Tooele County), Figure 12A (Uinta County), Figure 13A,B (Utah County), Figure 14B (Washington County), and Figure 15A,B (Weber County), indicates that the area is an oxidizer sink, meaning that the region removes oxidizers with increasing NOx concentrations and is VOC-limited. The discovery that these regions are oxidant sinks and VOC-limited is consistent with the findings of Womack et al. [30].

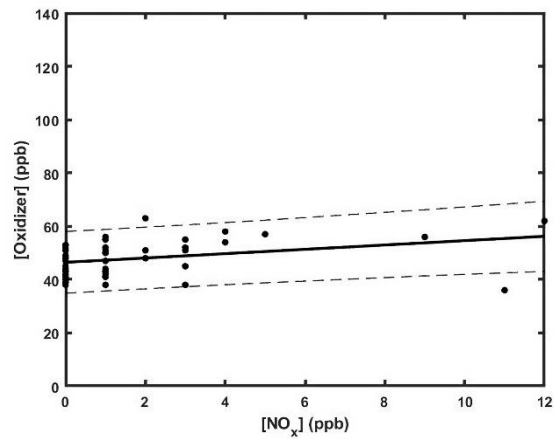

**Figure 2.** Box Elder County 2022 with 95% confidence interval to the fit shown by the dashed lines. [OX] = 0.81 [NO$_x$] + 46. The solid line represents the fit line for the linear regression and the dashed lines represent the 95% confidence interval in the fit.

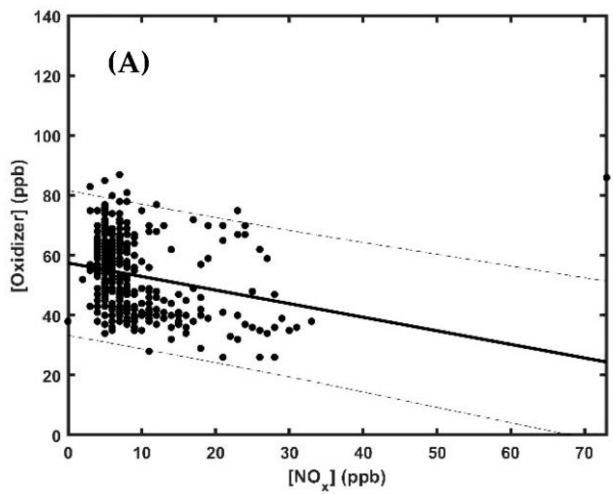 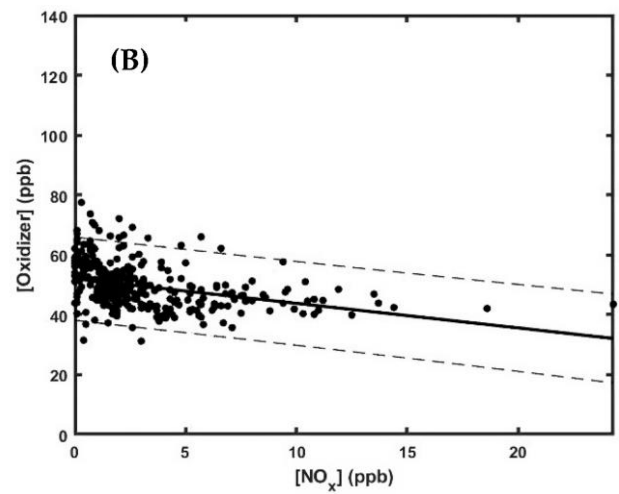

**Figure 3.** Cache County with 95% confidence interval to the fit shown by the dashed lines. (**A**)—2012 [OX] = −0.45 [NO$_x$] + 58; (**B**)—2022 [OX] = −0.82 [NO$_x$] + 52. The solid line represents the fit line for the linear regression and the dashed lines represent the 95% confidence interval in the fit.

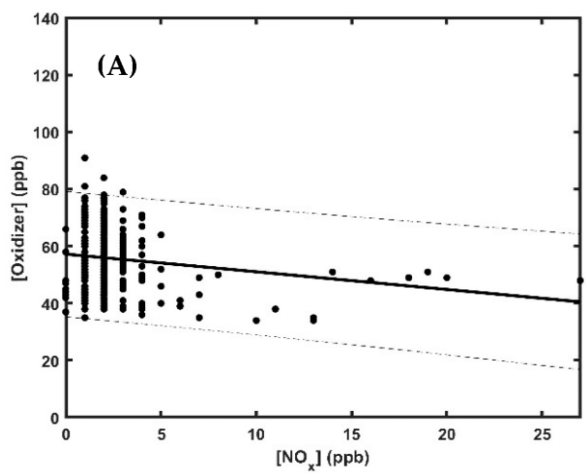 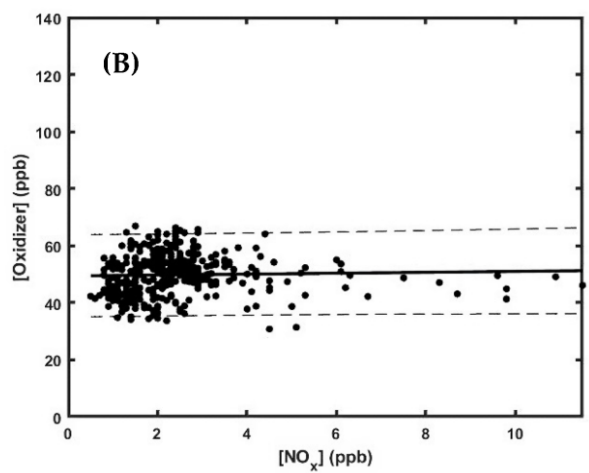

**Figure 4.** Carbon County with 95% confidence interval to the fit shown by the dashed lines. (**A**)—2012 [OX] = −0.62 [NO$_x$] + 57; (**B**)—2022 [OX] = 0.16 [NO$_x$] + 49. The solid line represents the fit line for the linear regression and the dashed lines represent the 95% confidence interval in the fit.

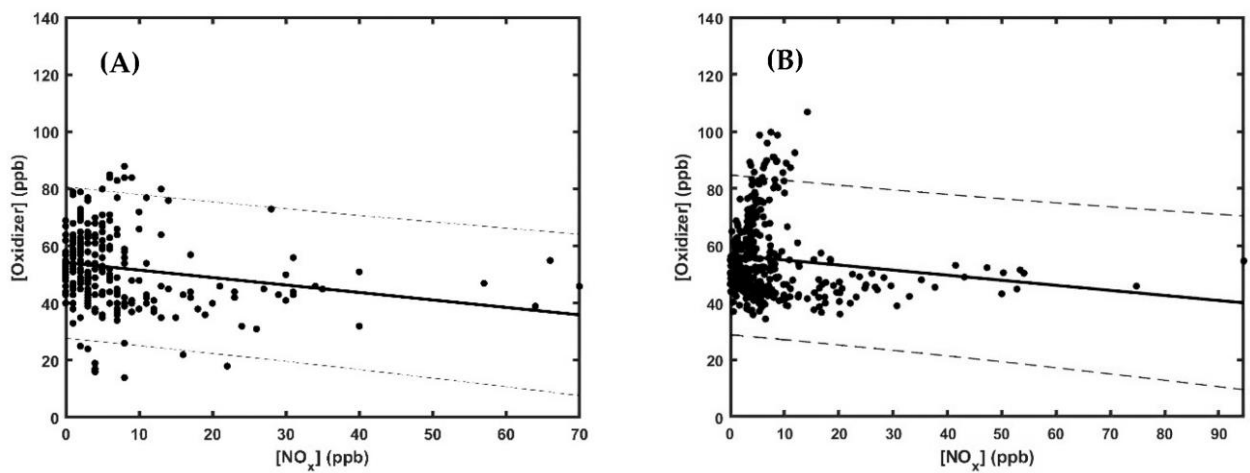

**Figure 5.** Davis County with 95% confidence interval to the fit shown by the dashed lines. (**A**)—2012 [OX] = −0.26 [NO$_x$] + 54; (**B**)—2022 [OX] = −0.18 [NO$_x$] + 57. The solid line represents the fit line for the linear regression and the dashed lines represent the 95% confidence interval in the fit.

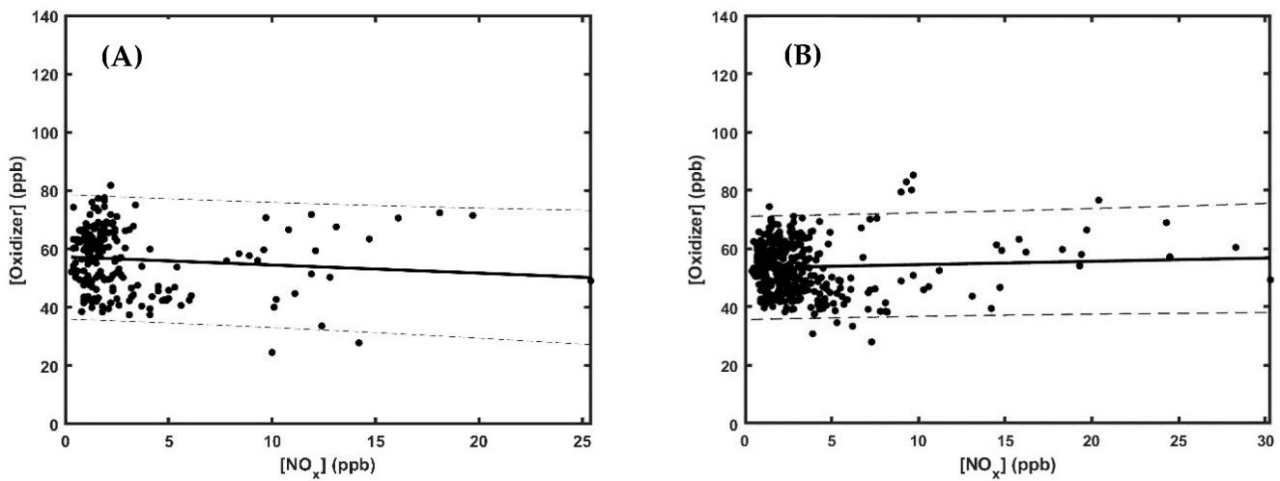

**Figure 6.** Duchesne County with 95% confidence interval to the fit shown by the dashed lines. (**A**)—2012 [OX] = −0.28 [NO$_x$] + 57; (**B**)—2022 [OX] = 0.11 [NO$_x$] + 53. The solid line represents the fit line for the linear regression and the dashed lines represent the 95% confidence interval in the fit.

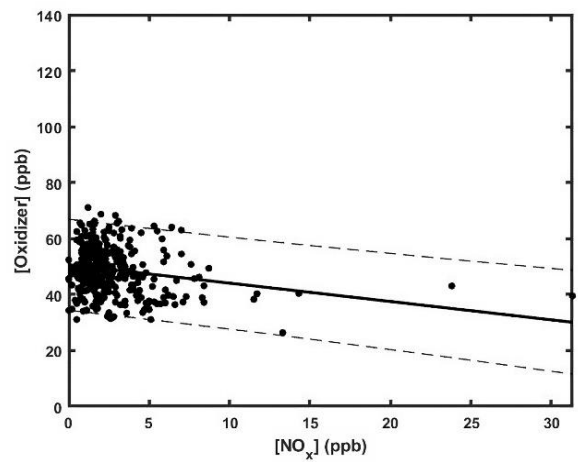

**Figure 7.** Iron County 2022 with 95% confidence interval to the fit shown by the dashed lines. [OX] = −0.66 [NO$_x$] + 51. The solid line represents the fit line for the linear regression and the dashed lines represent the 95% confidence interval in the fit.

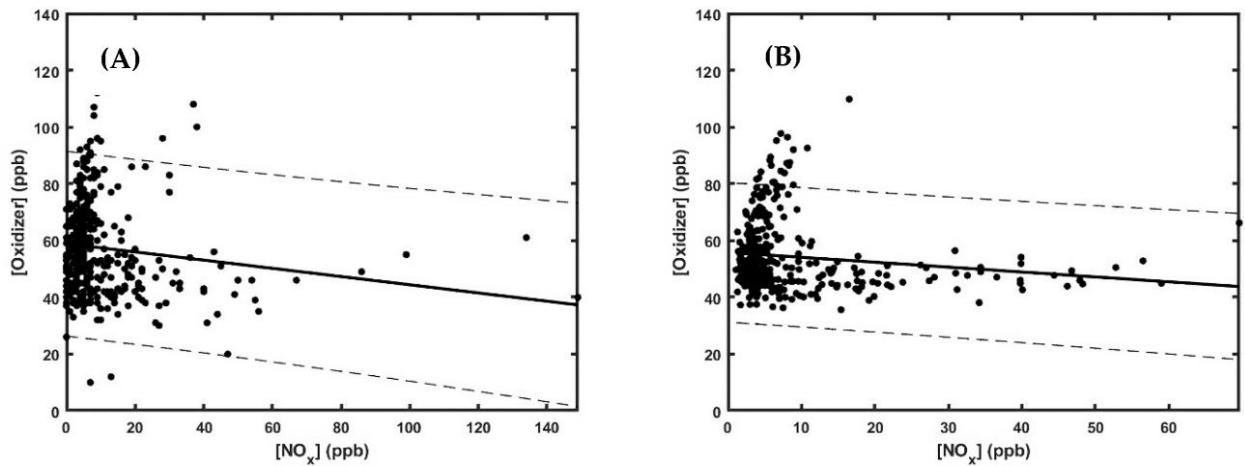

**Figure 8.** Salt Lake County with 95% confidence interval to the fit shown by the dashed lines. (**A**)—2012 [OX] = −0.14 [NO$_x$] + 59; (**B**)—2022 [OX] = −0.17 [NO$_x$] + 56. The solid line represents the fit line for the linear regression and the dashed lines represent the 95% confidence interval in the fit.

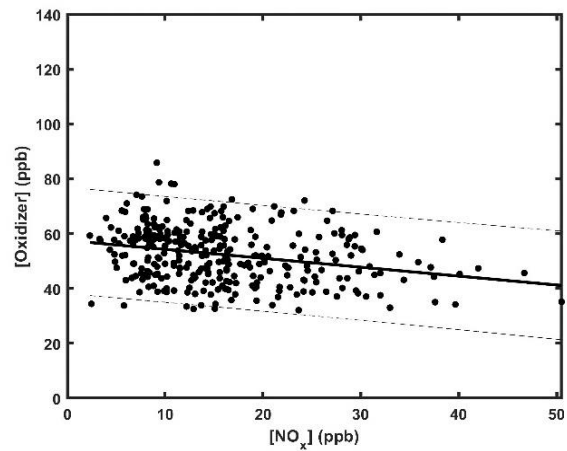

**Figure 9.** Sanpete with 95% confidence interval to the fit shown by the dashed lines. 2021 [OX] = −0.33 [NO$_x$] + 58. The solid line represents the fit line for the linear regression and the dashed lines represent the 95% confidence interval in the fit.

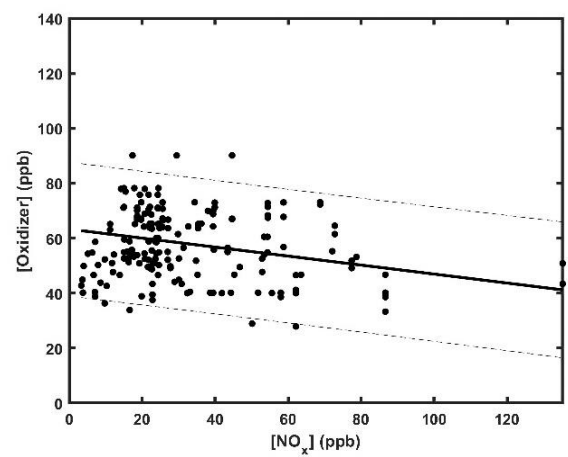

**Figure 10.** Sevier with 95% confidence interval to the fit shown by the dashed lines. 2017 [OX] = −0.16 [NO$_x$] + 64. The solid line represents the fit line for the linear regression and the dashed lines represent the 95% confidence interval in the fit.

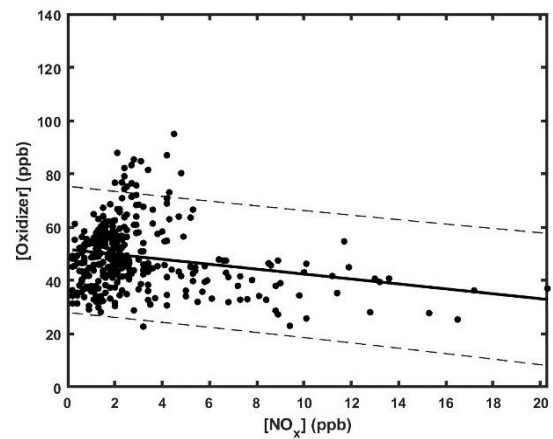

**Figure 11.** Tooele with 95% confidence interval to the fit shown by the dashed lines. 2022 [OX] = −0.92 [$NO_x$] + 52. The solid line represents the fit line for the linear regression and the dashed lines represent the 95% confidence interval in the fit.

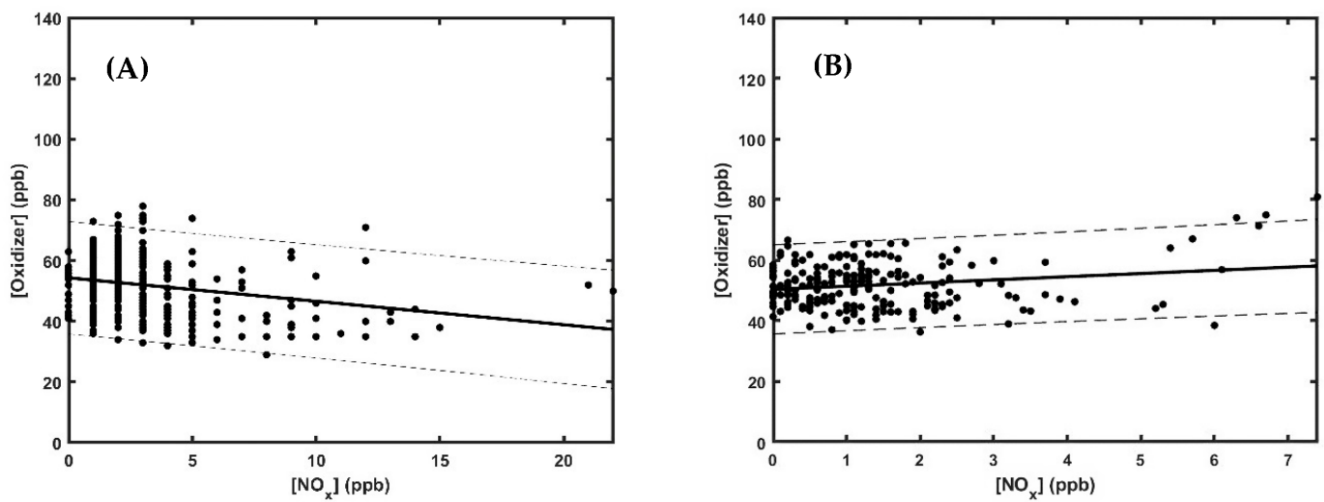

**Figure 12.** Uinta County with 95% confidence interval to the fit shown by the dashed lines. (**A**)—2012 [OX] = −0.77 [$NO_x$] + 54; (**B**)—2022 [OX] = 1.05 [$NO_x$] + 50. The solid line represents the fit line for the linear regression and the dashed lines represent the 95% confidence interval in the fit.

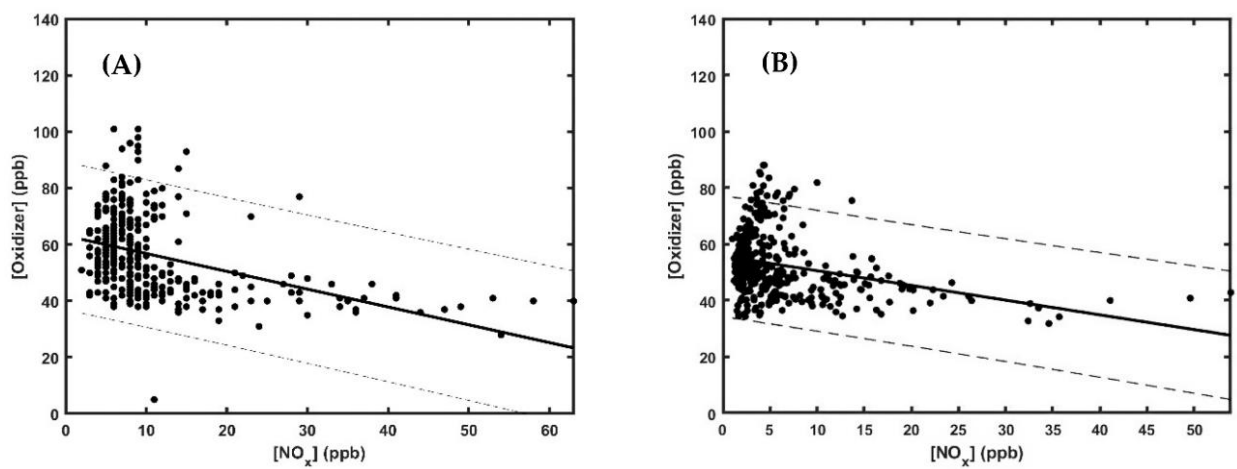

**Figure 13.** Utah County with 95% confidence interval to the fit shown by the dashed lines. (**A**)—2012 [OX] = −0.63 [$NO_x$] + 63; (**B**)—2022 [OX] = −0.52 [$NO_x$] + 56. The solid line represents the fit line for the linear regression and the dashed lines represent the 95% confidence interval in the fit.

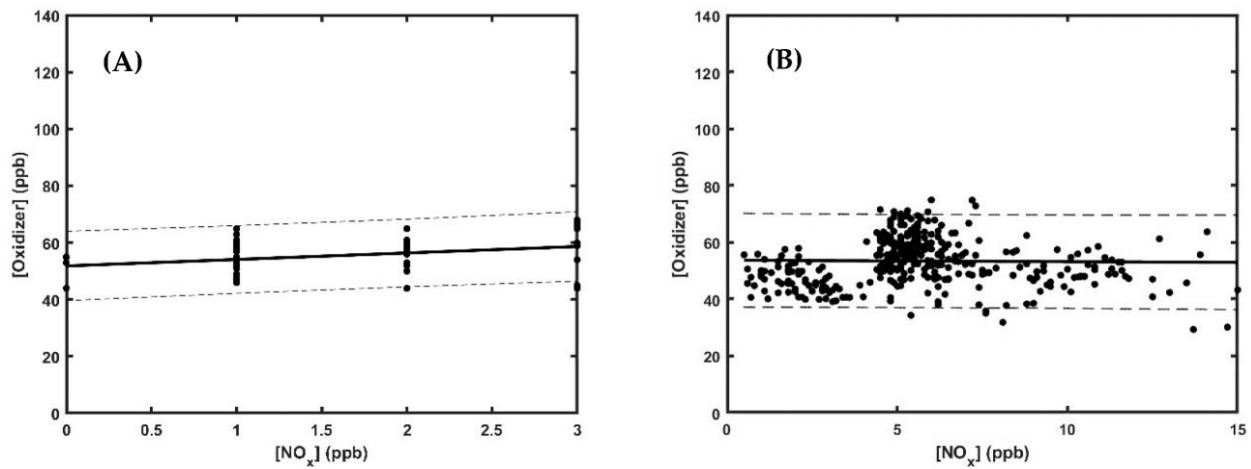

**Figure 14.** Washington County with 95% confidence interval to the fit shown by the dashed lines. (**A**)—2012 [OX] = 2.05 [NO$_x$] + 52; (**B**)—2022 [OX] = −0.05 [NO$_x$] + 54. The solid line represents the fit line for the linear regression and the dashed lines represent the 95% confidence interval in the fit.

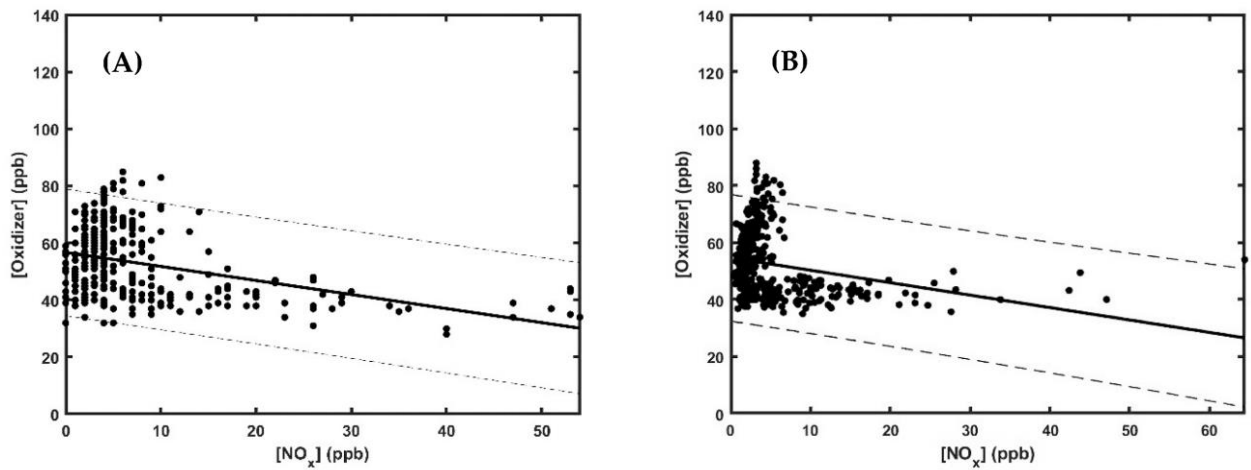

**Figure 15.** Weber County with 95% confidence interval to the fit shown by the dashed lines. (**A**)—2012 [OX] = −0.49 [NO$_x$] + 57; (**B**)—2022 [OX] = −0.44 [NO$_x$] + 55. The solid line represents the fit line for the linear regression and the dashed lines represent the 95% confidence interval in the fit.

The magnitude of the slope in this analysis indicates the local contribution of oxidizers to the overall concentration. If the magnitude of the slope increased over the 10-year period analyzed, it is concluded that the local contribution to the oxidizer concentration increased. Over the analyzed 10-year period, two counties were found to have increased their local contributions to oxidizer concentrations: Cache and Salt Lake. The opposite effect can be concluded if the magnitude of the slope decreased; the local contribution to oxidizer concentrations will have decreased. This occurred in Carbon, Davis, Duchesne, Uinta, Utah, Washington, and Weber counties. Interestingly, three counties show a change in the sign of the slope, which can be interpreted as transitioning from either a VOC- or NO$_x$-limited regime to the opposite regime over the ten-year period analyzed. The counties that have had a change of slope from either negative to positive include Carbon, Duchesne, and Uinta counties. This conclusion may also be applied to counties with slopes very close to zero, such as Salt Lake County. This means that the county is not clearly VOC- or NO$_x$-limited but rather in a transition state between the two regimes. As discussed later, this transition was observed in Carbon, Duchesne, and Uinta counties over the ten-year period analyzed but this transition from VOC- or NO$_x$-limited regimes can vary from month to month throughout the year.

All data included in the graphs are in units of parts per billion (ppb) and are plotted to adhere to a linear relationship (y = mx + b) with y being the oxidizer concentration, x being the $NO_x$ concentration, b being the oxidizer concentration in the absence of $NO_x$—or ozone concentrations—and m being the slope of the relationship between $NO_x$ and oxidizers with units of ppb oxidizer/ppb $NO_x$. The scatter in these plots represents the breakdown of the relationship described by reactions 1 and 2. This can be due to competing reactions, such as the formation of nitric acid from the reaction of $NO_2$ with OH, or due to meteorological effects, such as wind gusts, clouds, and dust storms that could hinder photolysis. When nitric acid is formed, $NO_x$ is removed from the cycle described by reactions 1 and 2 and the cycle needs to adjust before reaching equilibrium again. This may be more frequent during winter months when atmospheric conditions are more favorable for the formation of nitric acid that can react with $NH_3$ to produce ammonium-nitrate-nucleated particulate matter. On days with significant cloud cover and dust storm days, the rate of reaction 2 slows due to a decrease in the intensity of the photon flux needed to photolyze $NO_2$. This shifts the equilibrium of the cycle. Gusts of wind that do not accompany an increase in dust or clouds may still affect the mechanism described by reactions 1 and 2 by introducing more $NO_x$ and/or ozone, thereby causing the cycle to shift and re-equilibrate. The cycle described by reactions 1 and 2 will be ramping up or down most of the time due to these interferences and contribute to scatter in these plots. Only times when there are no interferences will the cycle be linear, and, as a result, data points will fall perfectly on the trendline.

Table 2 lists the change in slope for a given region between the years 2012 and 2022, and the regional and local contributions to $O_3$, which were determined from the intercept and slope of Figures 2–15. Included in Table 2 are indicators of anthropogenic influences on a given region, including population change, gross domestic product (GDP), and vehicle miles traveled (VMT) for each region studied. Local contributions to oxidizer concentrations are hypothesized to correlate with anthropogenic emissions in a region. As observed in Table 2, trends in local contributions to oxidant concentrations do not always follow indicators of anthropogenic activity. For some counties, there is an anti-correlation to these indicators that suggests other influences are more dominant. While the population change, GDP, and VMT increased in Salt Lake and Cache counties, these were not the counties with the largest increase in these indicators. Washington and Utah counties have the most significant increase in these indicators, yet both counties saw a decrease in local emissions. This same anti-correlation was seen in Davis, Weber, Uinta, and Duchesne counties. The population of Carbon County decreased, as did their local emissions.

Table 2. Summary of annual data for each county studied.

| County | Slope Change between 2012 to 2022) | Slope Change (%) | Absolute Change in Regional Contribution | Change in Regional Contribution (%) | Population Changes from 2012 to 2022 (%) [31] | GDP Changes from 2012 to 2021 (%) [32] | Vehicle Miles Travelled Changes from 2012 to 2021 (%) [33] | Annual Average Direction of Prevailing Winds [34] |
|---|---|---|---|---|---|---|---|---|
| Box Elder | - | - | - | - | 18.5 | 58.6 | 34.8 | - |
| Cache | Remained VOC | 82.22 | −9.58 | −5.51 | 20.5 | 74.8 | 25.8 | N-S |
| Carbon | VOC to $NO_x$ | −125.81 | −13.73 | −7.86 | −2.8 | 18.8 | 37.2 | N-S |
| Davis | Remained VOC | −30.77 | 4.57 | 2.48 | 17.0 | 61.3 | 21.3 | E-W |
| Duchesne | VOC to $NO_x$ | −139.29 | −7.02 | −4.03 | 3.6 | −7.8 | 46.8 | - |
| Iron | - | - | - | - | 34.5 | 94.6 | 45.2 | SSW-NNE |
| Salt Lake | Remained VOC | 21.43 | −5.20 | −3.06 | 13.3 | 72.4 | 14.9 | SSE-NNW |
| Sanpete | - | - | - | - | 7.3 | 60.6 | 44.5 | NNW-SSE |
| Sevier | - | - | - | - | 5.1 | 62.5 | 33.8 | NNE-SSW |
| Tooele | - | - | - | - | 29.0 | 32.6 | 19.4 | E-W |
| Uinta | VOC to $NO_x$ | −236.36 | −7.39 | −4.02 | 5.0 | −21.7 | 13.9 | W-E |
| Utah | Remained VOC | −17.46 | −11.71 | −7.4 | 30.2 | 110.8 | 33.3 | NW-SE |
| Washington | $NO_x$ to VOC | −102.44 | 2.36 | 1.24 | 35.7 | 119.0 | 58.0 | ENE-WSW |
| Weber | Remained VOC | −10.20 | −2.12 | −2.12 | 12.7 | 69.3 | 16.4 | S-N |

The intercept in each plot is a measurement of the regional transportation of ozone and its precursors into a county. This means that a certain amount of locally observed ozone is due to either regional transport or stratospheric intrusion and not the result of pollutants generated locally. In general, all counties have seen a slight decrease in the regional transport contribution to the oxidant concentration. Figure 16 shows the annual prevailing local wind directions at each airport in Utah. Overlaid on this map are the UDAQ permitted point source VOC and $NO_x$ emissions and their intensities. It should be noted that these are ground-level wind measurements taken at Utah's airports. This means that these ground-level measurements are affected by local change in topography (lake effects and mountain-valley flow) and often travel along the centerline of a north-south valley instead of in line with the surface wind direction at this latitude. The westerlies are the prevailing surface winds at this latitude and run from west to east across the state. Wind back trajectories were modeled using the National Oceanic and Atmospheric Administration (NOAA) HYSPLIT trajectory model for the Salt Lake Airport, Logan Airport in Cache County, Provo Airport in Utah County, and the St. George Airport in Washington County [35]. These wind back trajectories were modeled for both January (winter) and July (summer) at altitudes of 10 m and 1000 m above ground level and are seen in Figures S1–S8. It is seen that these winds often come over Southern California which is highly populated and known to have a lot of emissions. The calculated back wind trajectories typically enter Utah from California, but a few of the higher elevation back wind trajectories are calculated enter Northern Utah from over Oregon, Washington, and sometimes even Canada. This shows that the wind may be carrying emissions from other states into Utah which contributes to background ozone levels.

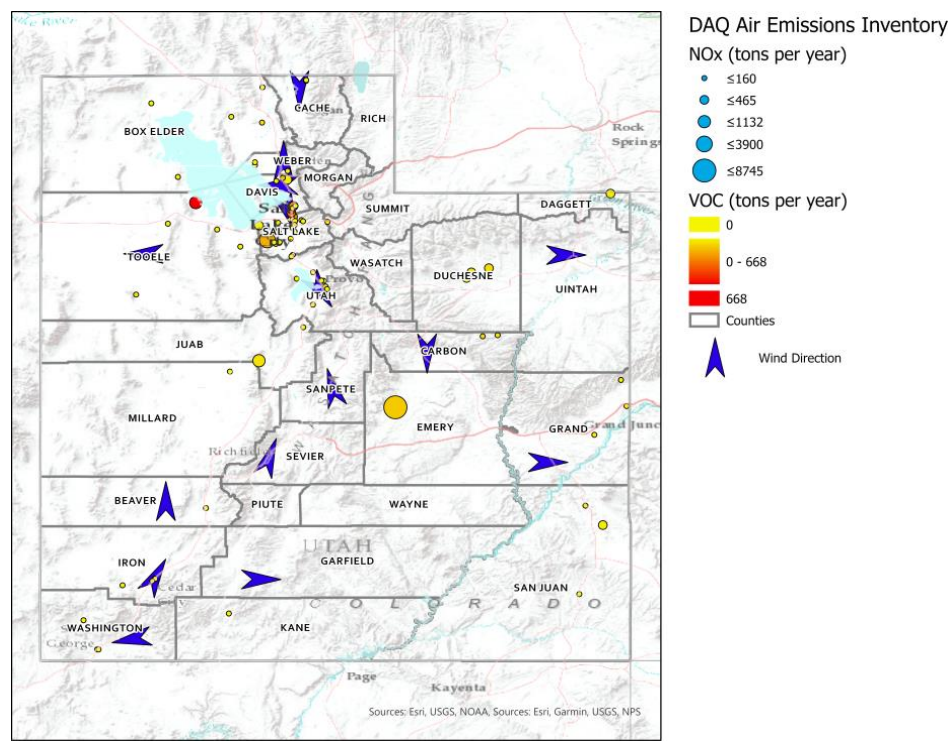

**Figure 16.** Utah, USA map of counties, UDAQ permitted point source emitters [36], and prevailing wind direction [34].

The UDAQ Air Emissions Inventory was used to identify point source of VOC and $NO_x$ emitters in the state of Utah (SGID 2023). The five largest industrial VOC emitters in Utah are all west of, or in Salt Lake and Davis Counties. The sum of VOC emissions of these five emitters produces 32% more emissions than the sum of the remaining 75 permitted emitters in Utah. The five largest emitters in this area are US Magnesium LLC—Rowley Plant just west of the Great Salt Lake in Tooele County, Big West Oil Refinery in Davis County, Chevron Products Company Salt Lake Refinery in Davis County, Holly Corporation HRMC and HEP Woods Cross operations in Davis County, and Tesoro Refining and Marketing Company LLC's Salt Lake City Refinery in Salt Lake County. These emitters are all west of Salt Lake City and the general wind direction is from west to east meaning that these emissions are blown into the Salt Lake Valley where the winds hit the mountains and are redirected either north or south along the Wasatch front. There are also many industrial $NO_x$ emitters in Salt Lake County. The top five $NO_x$ emitters produce 236% more $NO_x$ than the other 75 $NO_x$ emitters combined. These are PacifiCorp—Hunter Power Plant in Emery County, Kennecott Utah Copper LLC—Mine and Copperton Concentrator in Salt Lake County, US Magnesium LLC—Rowley Plant, Ash Grove Cement Company—Leamington Cement Plant in Millard County, and Mountain West Pipeline LLC—Kastler Marushack Compressor Station in Daggert County. US Magnesium and Kennecott Utah Copper will have their $NO_x$ emissions blown west to the Salt Lake Valley to mix with the VOC emissions in the area producing ozone.

In general, there is a northernly wind that flows north along the mountain range from Iron County through Beaver, Sevier, and Sanpete into Utah County. This means that emissions from these counties can be transported into Utah county. The prevailing wind also blows emissions from Salt Lake County into Utah county. While winds in Utah can move emissions from county to county, background ozone levels across the fourteen sampled counties are in the region of 50–57 ppb meaning that some emissions may be coming from other states or countries. It is significant that regional contributions to oxidant concentrations were smaller by 3–8 ppb between the years 2012 to 2022.

Table 3 provides monthly averaged ground level wind patterns. These data aids in understanding the movement of emissions within Utah on a monthly basis. The largest effect is the trend that shows the prevailing wind hitting the Wasatch mountains and then either being pushed north from Washington County through Iron, Sevier, and Sanpete counties to Utah County or south from Davis County through Salt Lake County to Utah County. While Utah County does not have any major VOC or $NO_x$ source emitters, it still suffers from elevated levels of ozone. This may be from county-to-county transport of precursor species meeting in Utah County. Emissions that would be transported from county-to-county may be $NO_x$ emissions from vehicles and biogenics. The I-15 interstate runs along the mountains in the described path and may contribute to elevated $NO_x$ emissions. At Utah Department of Transportation (UDOT) station 011-0020 (SR 68 500 South Bountiful in on the border of Salt Lake and Davis Counties) an annual average of daily traffic of 157,910 vehicles are counted (Transportation 2023). Of these vehicles, 16.1%, or 25,424 vehicles are trucks. This high volume of vehicles is responsible for large amounts of $NO_x$ emissions which can be transported from county-to-county. The counties south of Utah -County are considered more rural and may have higher biogenic emissions which are VOC emissions that will contribute to summer spikes in VOC concentrations.

One prominent regional transport mechanism is the wind channel blowing from California and Nevada through Tooele County, following the west-to-east surface winds, into the Wasatch front. Evidence of state-to-state pollution transfer has been studied and shows that pollution from wildfires and emissions of states west of Utah, such as California and Nevada, are partially responsible for adding to the regional contributions to the oxidizer concentrations in Utah [15,37–39]. There are, however, environmental impacts to the oxidizer concentrations in Utah that do not originate in the United States. In the Utah Division of Air Quality Demonstration from 2021, UDAQ quoted data generated in a study by Ramboll Inc. (Arlington, VA, USA) who modeled the contributions to ozone at the Bountiful, UT EPA site (Quality 2021). Here they concluded that only 15% of ozone in the area was from Utah anthropogenic emissions while 10% was from US anthropogenic emissions, 56% from global natural and re-circulated sources, and 19% from non-US anthropogenic emissions with a possible ±4% error. They completed the same modeling studies for other sites around Utah and generated similar findings.

Table 3. Origin of Prevailing Wind in Each County by Month [34].

| County | Jan | Feb | Mar | Apr | May | Jun | Jul | Aug | Sep | Oct | Nov | Dec | Annual |
|---|---|---|---|---|---|---|---|---|---|---|---|---|---|
| Box Elder | - | - | - | - | - | - | - | - | - | - | - | - | - |
| Cache | N | N | N | N | N | N | N | S | N | N | N | N | N |
| Carbon | N | N | N | N | N | N | N | N | N | N | N | N | N |
| Davis | E | E | E | E | E | E | E | E | E | E | E | E | E |
| Duchesne | - | - | - | - | - | - | - | - | - | - | - | - | - |
| Iron | SSW | SW | SSW | SSW | SSW | SSW | SW | SSW | SSW | SW | N | SSW | SSW |
| Salt Lake | S | S | SSE | SSE | SSE | S | SSE | SSE | SSE | SE | SE | S | SSE |
| Sanpete | SE | N | SE | SW | S | S | SE | S | SE | SE | SE | SE | SE |
| Sevier | SW | NE | NW | SW | W | SW | W | NW | NW | NW | SW | W | SW |
| Tooele | NW | NW | E | NW | E | E | E | E | E | E | E | E | E |
| Uinta | W | W | WNW | W | W | W | W | W | W | W | WNW | W | W |
| Utah | NW | NW | NW | NW | NW | NW | SE | SE | SE | SE | SSE | SSE | NW |
| Washington | E | ENE | ENE | W | W | W | W | ENE | ENE | ENE | E | E | ENE |
| Weber | SSE | S | SSE | S | S | S | S | S | S | S | S | S | S |

　　In the past, UDAQ has said that a semi-permanent low-pressure system exists off the coast of China. This allows for pollutants to be pushed to the upper troposphere where winds from the region transport pollutants towards the western United States which arrive on the western shores of the US within days to weeks. Ozone that exists at these altitudes exists in an environment that has a "relative lack of chemical sinks" and low temperatures [40]. In contrast to the low-pressure system off the Chinese coast, the US coast has a semi-permanent high-pressure system which brings down the air from the upper troposphere to surface level over the western United States. This pattern can be seen in Figure 17. Topography of this mountainous region creates $O_3$ mixing down the slopes of the mountains which leads to high altitude areas experiencing impacts from intercontinental transport of $O_3$ compared to lower-altitude areas. UDAQ states, "This intercontinental transport persists throughout the summer season in Utah, leading to enhancements of local ozone concentrations" [40]. UDAQ has noted on a separate occasion that 20% of the West's total ozone concentrations are due to intercontinental transport of both ozone and $NO_x$ species and that this percentage grows every year 41. NASA's Jet Propulsion Laboratory (JPL) found that while the production of ozone-forming pollutants was reduced by 21% between 2005 and 2010 the concentrations of ozone in the region did not drop [8]. JPL concluded that this was due to naturally occurring processes and from transport of pollutants across the Pacific Ocean. One of these naturally occurring processes is the contribution from the stratosphere due to the "natural up-and-down cycle of the upper-atmosphere winds". UDAQ estimates that stratospheric intrusions can attribute 20–40 ppb of ozone to the ozone concentrations at ground level in Utah [41].

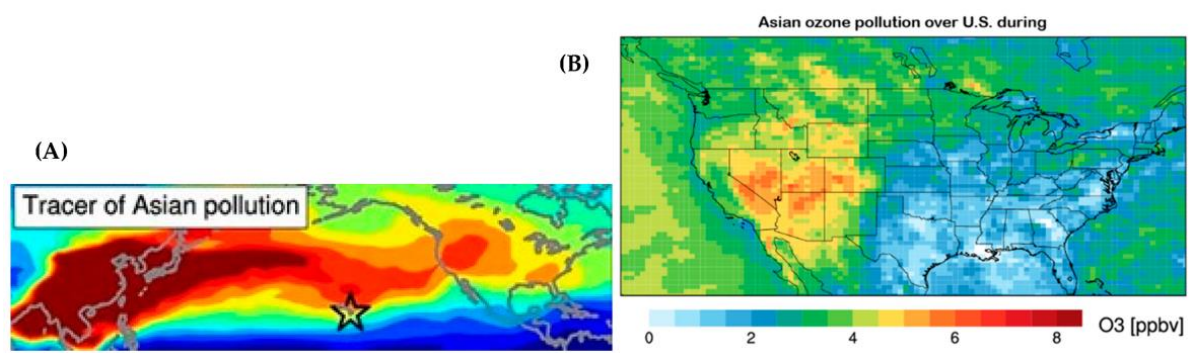

Figure 17. (**A**) shows typical transport of pollution from Asia to the United States. The star indicates where Hawaii is situated. (**B**) shows Asian ozone pollution distributed over the US. Images used with permission from UDAQ [42,43].

3.1.2. Monthly Data for Each County

Figures 18–20 show the monthly averaged regional and local contributions to oxidant concentrations for the years 2012 and 2022 for the 14 counties studied in this work. Plots of these data are found in Figures S9–S19. The shape of these plots shows how the regional and local contributions to oxidant concentrations vary throughout a year for each county. Evaluation of the pattern in these figures shows that most counties are in a transitional state for the winter months of November, December, January, and February. This can be seen in the regional contribution to oxidant figures (Figure 18B,D, Figure 19B,D and Figure 20B,D) as the regional contribution to oxidant concentrations during winter months is close to zero. There is a general rise in both regional and local contributions to oxidant concentrations during the summer months which is attributed to an increase in biogenic emissions. In Utah, groundcover snow and cold temperatures prevent biogenic emissions during winter months. Biogenic emissions add to the total VOC concentrations in a region leaving areas more $NO_x$-limited in summer months. One county which does not follow this trend is Uinta County which has a large regional contribution to its oxidant concentration occurring in March. The local contribution to oxidants trend shows that generally during summer months, regions are more $NO_x$-limited which is consistent with areas being impacted by biogenic emissions that increase VOC concentrations.

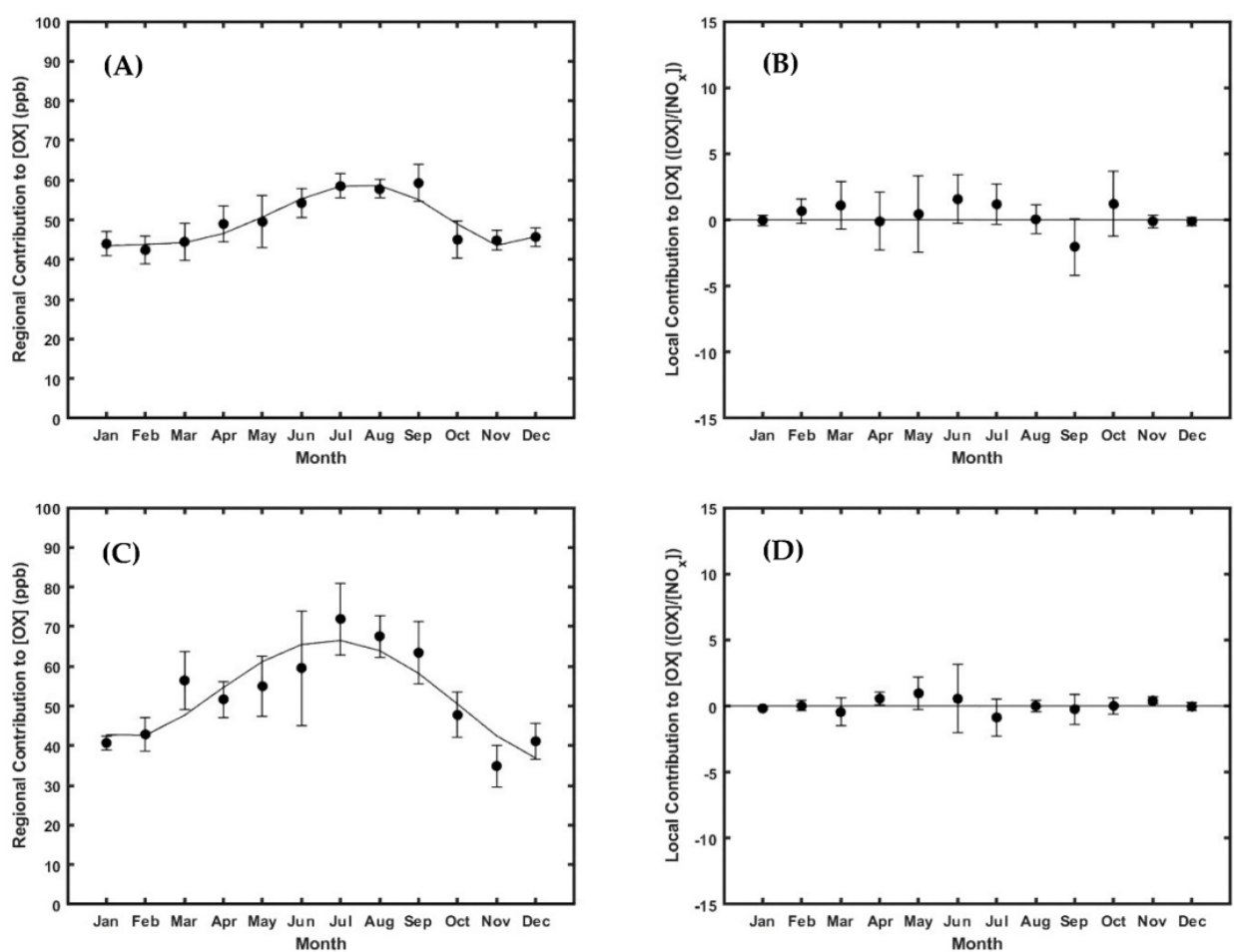

**Figure 18.** Cache County monthly analysis with 95% confidence interval to the fit. (**A**)—Regional contribution to the oxidant concentration, 2022; (**B**)—Local contribution to the oxidant concentration, 2022; (**C**)—Regional contribution to the oxidant concentration, 2012; (**D**)—Local contribution to the oxidant concentration, 2012.

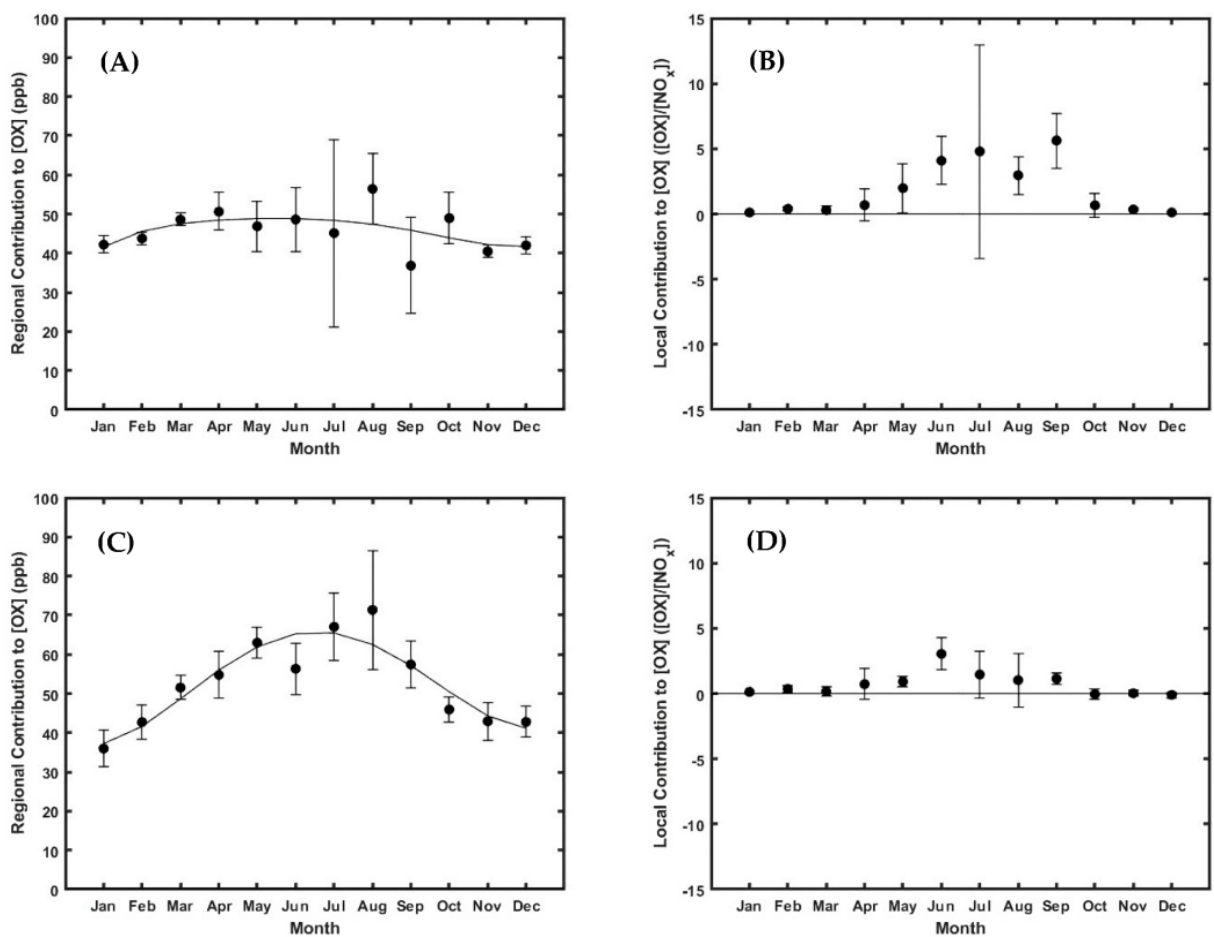

**Figure 19.** Salt Lake County monthly analysis with 95% confidence interval to the fit. (**A**)—Regional contribution to the oxidant concentration, 2022; (**B**)—Local contribution to the oxidant concentration, 2022; (**C**)—Regional contribution to the oxidant concentration, 2012; (**D**)—Local contribution to the oxidant concentration, 2012.

The sign and magnitude of the slopes, interpreted as the local contribution to oxidant concentrations in Figures 18–20 plots can be used to identify if counties are transitional or if they belong to either a $NO_x$ or VOC-limited regime. Three of these counties, Iron, Sanpete, and Sevier, are in a transitional regime between $NO_x$ and VOC-limited as the local contribution to oxidant concentrations for these counties remain very close to zero throughout the year. These assumptions, however, do not include Tooele County which is VOC-limited when evaluated using annual averaged data but on a monthly averaged basis can be $NO_x$-limited.

3.1.3. Vehicle Miles Traveled

Figure 21 shows modeled vehicle miles traveled (VMT) per county for each month of the year. Modeling was done using the EPA's Motor Vehicle Emission Simulator (MOVES v3.0) to estimate the number of VMT per county. This shows that there is a rise in vehicle miles traveled in every county over the summer months with August generally having the most VMT and February having the least. This pattern can be seen in the local contribution to oxidant plots in Figure 18, Figure 19, Figure 20 and Figures S9–S19. This pattern can also be seen in the regional contribution to oxidant plots in Figure 18, Figure 19, Figure 20 and Figures S9–S19 showing that this trend may exist in counties contributing to regional concentrations of oxidants.

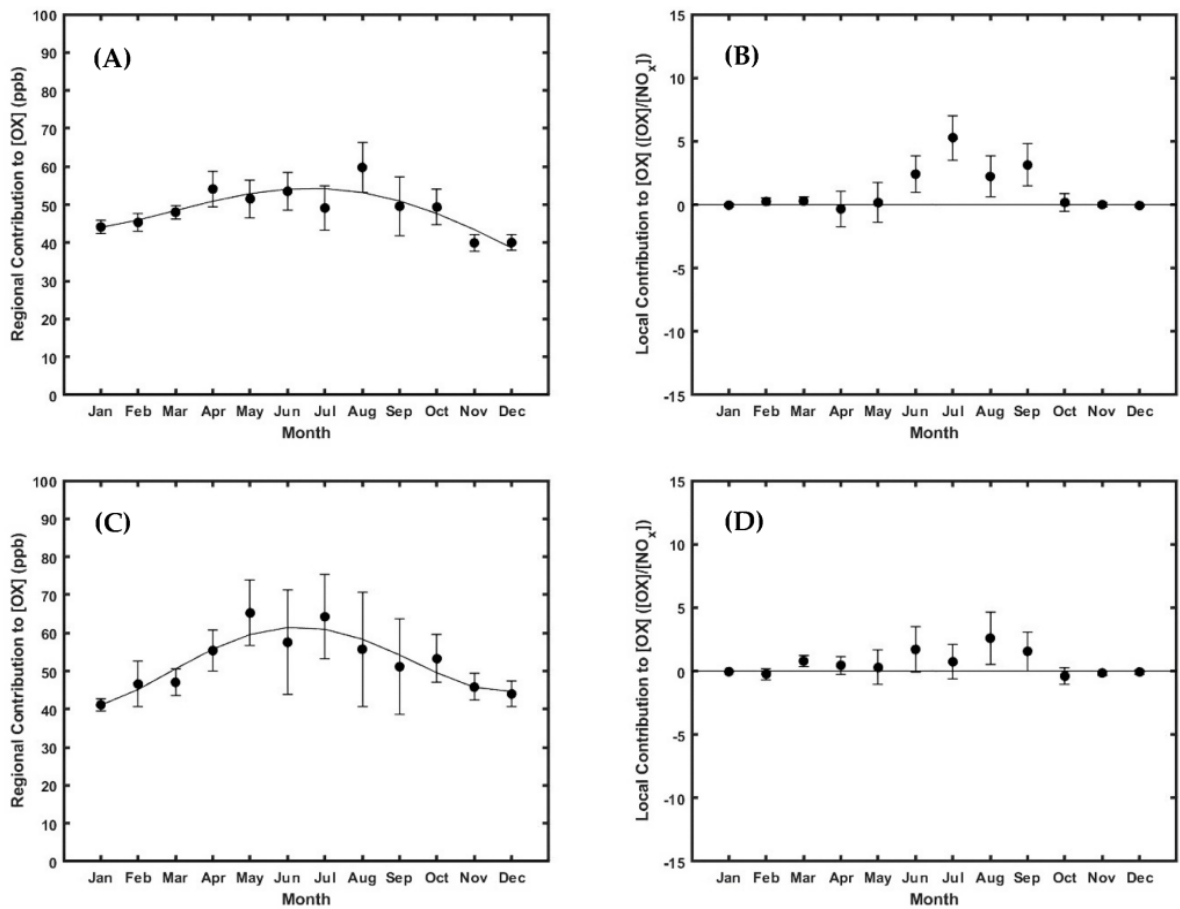

**Figure 20.** Utah County monthly analysis with 95% confidence interval to the fit. (**A**)—Regional contribution to the oxidant concentration, 2022; (**B**)—Local contribution to the oxidant concentration, 2022; (**C**)—Regional contribution to the oxidant concentration, 2012; (**D**)—Local contribution to the oxidant concentration, 2012.

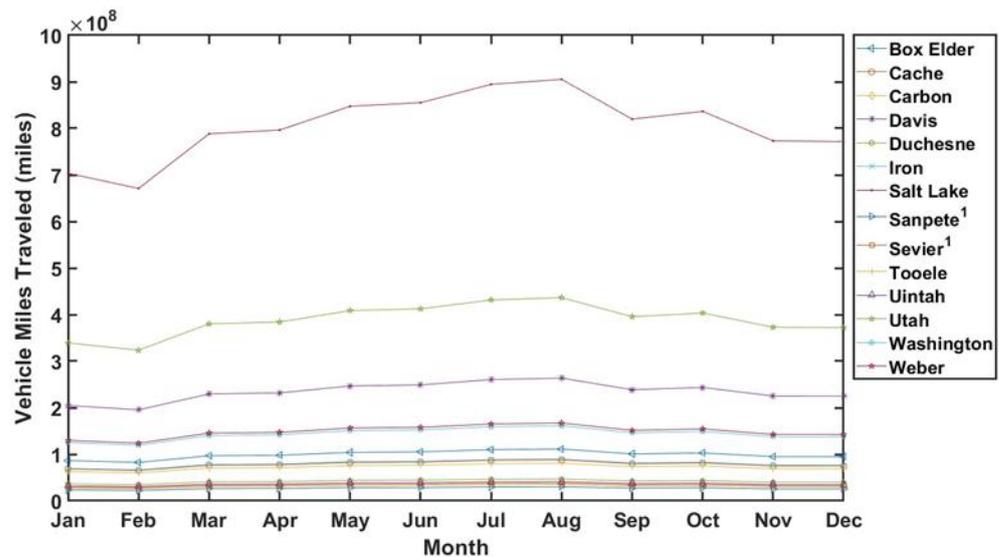

**Figure 21.** Modeled MOVES, VMT for each county by month for the year 2022. [1]-Sanpete data from year 2021 and Sevier data from year 2017.

### 3.1.4. Local Compared to Regional Contributions to Oxidants

The general trend in the regional contributions to oxidants in each county between the years 2012 and 2022 is similar, often with similar magnitudes as well. From this we can conclude that regional contributions to oxidizer concentrations remain mostly constant between the years 2012 and 2022. The general trend in local contributions to oxidants for each county is also similar between the years 2012 and 2022, however, in some counties, such as Cache, Davis, Salt Lake, Utah, and Weber, the trend has been shifted up on the *y*-axis. This means that there has been an increase in local contributions to oxidizer concentrations in these counties, which is expected in accordance with the increase in population, VMT, and GDP over the 10-year period studied.

*3.2. Fire Year (2021)*

Figures 22–24 show plots of oxidants versus $NO_x$ for three counties (Cache, Salt Lake, and Utah) that were heavily impacted by smoke from wildfires in the Pacific Northwest during the summer months of 2021. It is known that wildfire smoke increases the regional contribution of emissions in a region [44–47]. We are interested in the increase of ozone, VOCs, and $NO_x$ that accompany wildfire-smoke-impacted regions. This means that an increase in regional contributions to oxidants coupled with a decrease in local contribution to oxidant concentrations should be observed in counties impacted by wildfire smoke. This relationship is observed in all three counties. In Cache County, in June 2021, the regional contribution to oxidant concentrations is 50 ppb. In August 2021, the regional contribution to oxidant concentrations rises to 70 ppb, a 40% increase, while the local contribution to oxidant concentration in June 2021 is 4 $[OX]/[NO_x]$, and drops to 1 $[OX]/[NO_x]$ in August 2021.

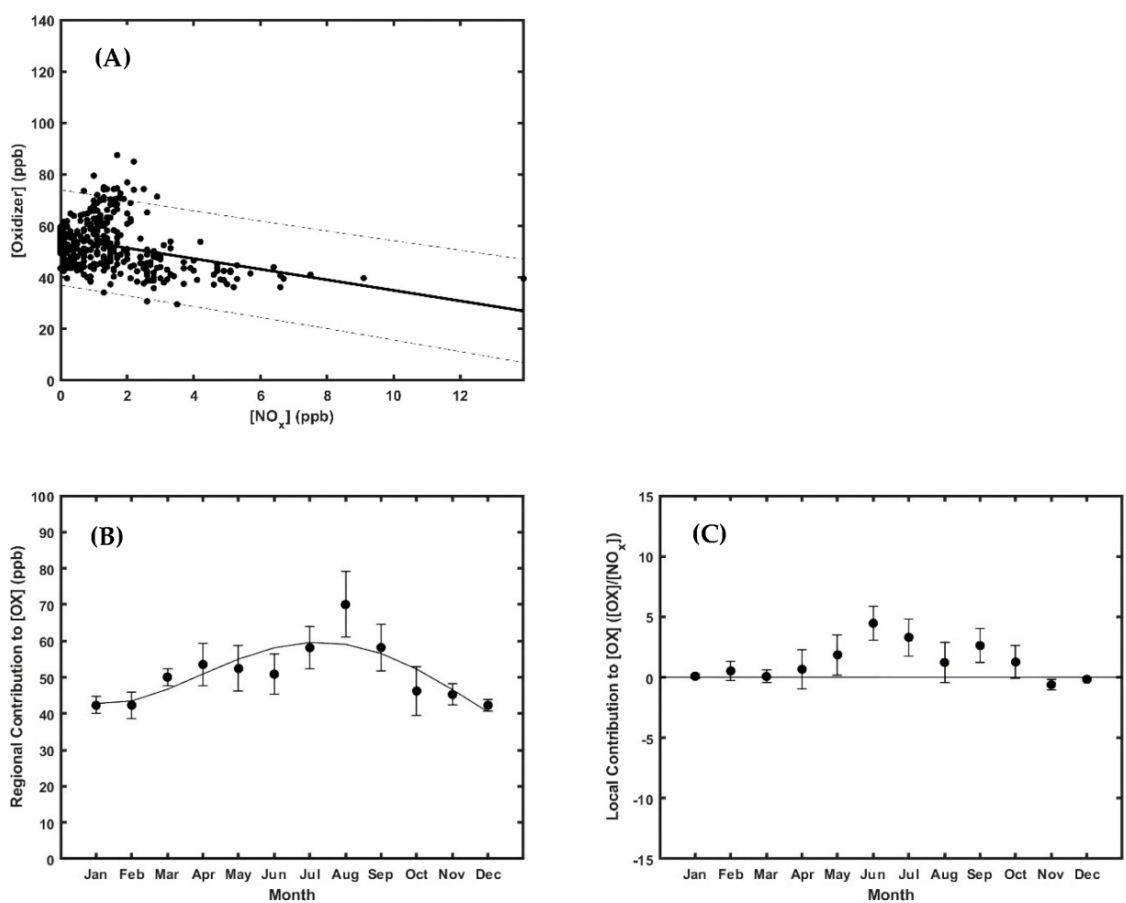

**Figure 22.** Cache County for 2021 with 95% confidence interval to the fit. (**A**)—Annual 2021 fit $[OX] = -2.06\,[NO_x] + 55.58$; (**B**)—Monthly 2021 Intercept; (**C**)—Monthly 2021 Slope.

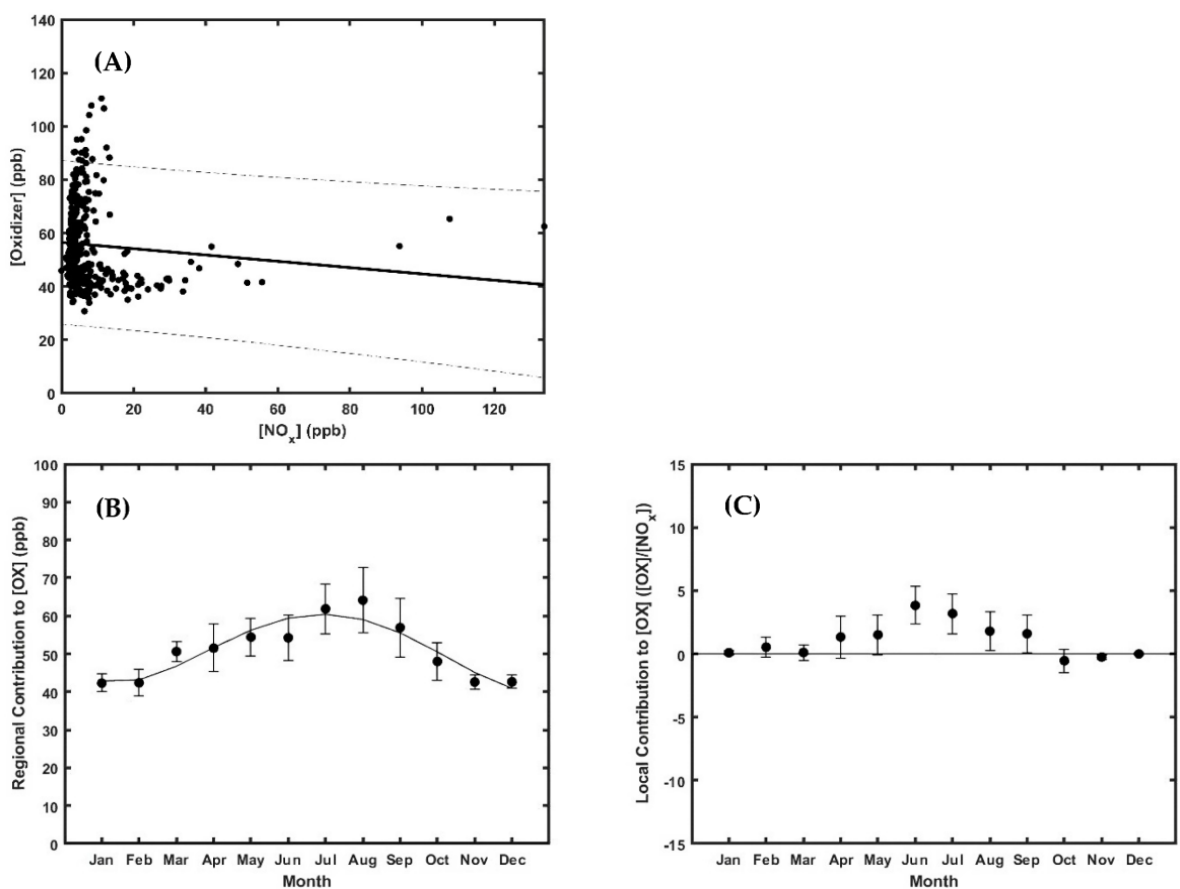

**Figure 23.** Salt Lake County for 2021 with 95% confidence interval to the fit. (**A**)—Annual 2021 fit [OX] = −0.12 [NO$_x$] + 56.6; (**B**)—Monthly 2021 Intercept; (**C**)—Monthly 2021 Slope.

In Salt Lake County, the change in the regional contribution to oxidants is not as dramatic as what is observed in Cache County. In June 2021, the regional contribution to oxidant concentrations is 54 ppb. In August 2021, the regional contribution to oxidant concentrations increases to 64 ppb, an 18% increase, while the local contribution to oxidant concentration in June 2021 is 4 [OX]/[NO$_x$], and drops to 2 [OX]/[NO$_x$] in August 2021.

In Utah County, in June 2021, the regional contribution to oxidant concentrations is 54 ppb. In August 2021, the regional contribution to oxidant concentrations increases to 64 ppb, an 18% increase, while the local contribution to oxidant concentration in June 2021 is 4 [OX]/[NO$_x$], and drops to 2 [OX]/[NO$_x$] in August 2021.

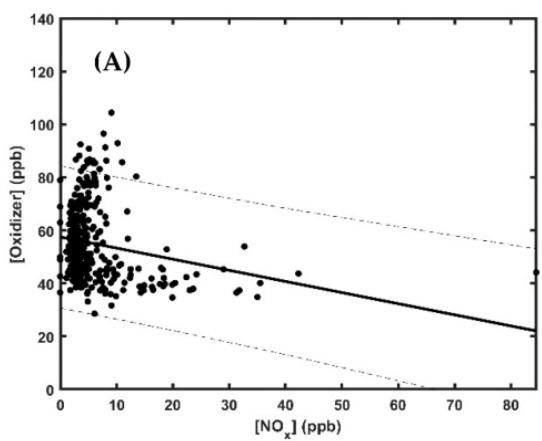

**Figure 24.** *Cont.*

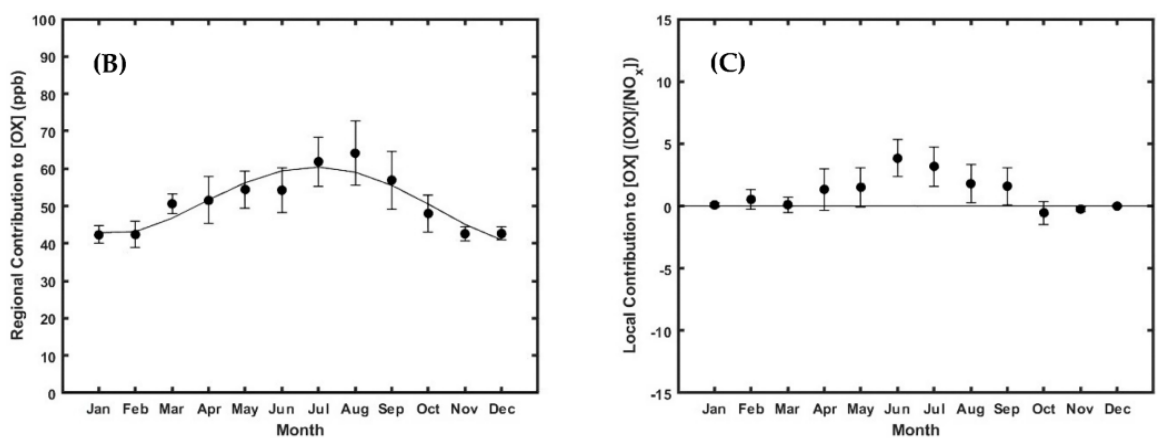

**Figure 24.** Utah County for 2021 with 95% confidence interval to the fit. (**A**)—Annual 2021 fit $[OX] = -0.41\ [NO_x] + 57.57$; (**B**)—Monthly 2021 Intercept; (**C**)—Monthly 2021 Slope.

## 4. Conclusions

The method of performing a linear regression on oxidizer versus $NO_x$ concentrations to deconvolve the influence of regional and local contributions of oxidants on a region has been used to understand the oxidizing power in the atmosphere. Conclusions about whether a region is $NO_x$- or VOC-limited are also able to be made from this analysis. The analysis of these data also provides information about the regional versus local contributions to oxidants on a county basis. This information can help stakeholders, legislation, and policy make informed decisions to guide future mitigation strategies as to the reduction of which precursor, $NO_x$ or VOCs, is most effective in reducing ozone production without the expenditure of an air sampling campaign. While this method does have limitations, it is an important starting point for understanding the oxidative nature of the atmosphere in Utah using currently employed infrastructure.

The Wasatch Front in Utah is currently a non-attainment region for ozone. It was found that the background concentrations of ozone range from 46–64 ppb and the current EPA standard for ozone is 70 ppb. This means that in some counties in Utah, attainment of the EPA NAAQS standard for ozone may be a significant challenge as the vast majority of ozone found in some counties does not originate in either the county or the state of Utah.

All counties, except for Washington County, were found to be VOC-limited in the year 2012. This shifted to more counties being in a transitional state (Carbon, Duchesne, Uinta, Salt Lake) or $NO_x$-limited (Box Elder, Cache, Duchesne, Uinta) in the year 2022. Local contributions to ozone have increased in Cache and Salt Lake between the years 2012 and 2022, but decreased in Carbon, Davis, Duchesne, Uinta, Utah, Washington, and Weber.

Regional contributions to oxidant concentrations were seen to decrease slightly; however, the decrease in some counties may be hidden in the uncertainties of these values. Regional contributions and local contributions to oxidant concentrations spike in the summer months and, in many cases, local contributions to oxidants exist in a transitional state between being $NO_x$- and VOC-limited for the winter months of November through February. This summertime spike in local contributions to oxidants could, in part, be due to biogenic emissions in the summertime that do not exist in the wintertime. These spikes also track the general trend of vehicle miles travelled in Utah during those months.

Smoke from wildfires has been found to increase the regional contribution to oxidants and shift the local regime to be more $NO_x$-limited by the influx of VOCs into a region.

**Supplementary Materials:** The following supporting information can be downloaded at: https://www.mdpi.com/article/10.3390/atmos14081262/s1, Figure S1: Logan (Cache County) wind back trajectory at 10 m above ground level at 14:00 local time. (A)—1 January 2022 (B)—1 July 2022; Figure S2: Salt Lake City (Salt Lake County) wind back trajectory at 10 m above ground level at 14:00 local time. (A)—1 January 2022 (B)—1 July 2022; Figure S3: Provo (Utah County) wind back trajectory at 10 m above ground level at 14:00 local time. (A)—1 January 2022 (B)—1 July 2022; Figure S4: St. George (Washington County) wind back trajectory at 10 m above ground level at 14:00 local time. (A)—1 January 2022 (B)—1 July 2022; Figure S5: Logan (Cache County) wind back trajectory at 1000 m above ground level at 14:00 local time. (A)—1 January 2022 (B)—1 July 2022; Figure S6: Salt Lake City (Salt Lake County) wind back trajectory at 1000 m above ground level at 14:00 local time. (A)—1 January 2022 (B)—1 July 2022; Figure S7: Provo (Utah County) wind back trajectory at 1000 m above ground level at 14:00 local time. (A)—1 January 2022 (B)—1 July 2022; Figure S8: St. George (Washington County) wind back trajectory at 1000 m above ground level at 14:00 local time. (A)—1 January 2022 (B)—1 July 2022; Figure S9: Box Elder County monthly analysis with 95% confidence interval to the fit. (A)—Regional contribution to the oxidant concentration, 2022 (B)—Local contribution to the oxidant concentration, 2022; Figure S10: Carbon County monthly analysis with 95% confidence interval to the fit. (A)—Regional contribution to the oxidant concentration, 2022 (B)—Local contribution to the oxidant concentration, 2022 (C)—Regional contribution to the oxidant concentration, 2012 (D)—Local contribution to the oxidant concentration, 2012; Figure S11: Davis County monthly analysis with 95% confidence interval to the fit. (A)—Regional contribution to the oxidant concentration, 2022 (B)—Local contribution to the oxidant concentration, 2022 (C)—Regional contribution to the oxidant concentration, 2012 (D)—Local contribution to the oxidant concentration, 2012; Figure S12: Duchesne County monthly analysis with 95% confidence interval to the fit. (A)—Regional contribution to the oxidant concentration, 2022 (B)—Local contribution to the oxidant concentration, 2022 (C)—Regional contribution to the oxidant concentration, 2012 (D)—Local contribution to the oxidant concentration, 2012; Figure S13: Iron County monthly analysis with 95% confidence interval to the fit. (A)—Regional contribution to the oxidant concentration, 2022 (B)—Local contribution to the oxidant concentration, 2022; Figure S14: Sanpete monthly analysis with 95% confidence interval to the fit. (A)—Regional contribution to the oxidant concentration, 2021 (B)—Local contribution to the oxidant concentration, 2021; Figure S15: Sevier County monthly analysis with 95% confidence interval to the fit. (A)—Regional contribution to the oxidant concentration, 2017 (B)—Local contribution to the oxidant concentration, 2017; Figure S16: Tooele County monthly analysis with 95% confidence interval to the fit. (A)—Regional contribution to the oxidant concentration, 2022 (B)—Local contribution to the oxidant concentration, 2022; Figure S17: Uinta County monthly analysis with 95% confidence interval to the fit. (A)—Regional contribution to the oxidant concentration, 2022 (B)—Local contribution to the oxidant concentration, 2022 (C)—Regional contribution to the oxidant concentration, 2012 (D)—Local contribution to the oxidant concentration, 2012; Figure S18: Washington County monthly analysis with 95% confidence interval to the fit. (A)—Regional contribution to the oxidant concentration, 2022 (B)—Local contribution to the oxidant concentration, 2022 (C)—Regional contribution to the oxidant concentration, 2012 (D)—Local contribution to the oxidant concentration, 2012; Figure S19: Weber County monthly analysis with 95% confidence interval to the fit. (A)—Regional contribution to the oxidant concentration, 2022 (B)—Local contribution to the oxidant concentration, 2022 (C)—Regional contribution to the oxidant concentration, 2012 (D)—Local contribution to the oxidant concentration, 2012.

**Author Contributions:** C.E.F.: Writing of the original draft and data analysis. R.T.: Writing review and editing, aided in data analysis and funding acquisition. J.C.H.: Conceptualization, writing review and editing, supervision, funding acquisition, and aided in analysis. All authors have read and agreed to the published version of the manuscript.

**Funding:** This work was supported by the National Science Foundation, grant #2114655.

**Data Availability Statement:** Available on BYU Scholar's archive https://scholarsarchive.byu.edu/data/52 (accessed on 17 May 2023).

**Conflicts of Interest:** The authors declare no conflict of interest.